\documentclass[pra,aps,amssymb,showpacs,floatfix]{revtex4}
\usepackage[dvips]{graphicx}

\newcommand{\size}{12}
\begin{document}



\title{Floquet Formalism of Quantum Pumps}

\author{Sang Wook Kim}
\address{Department of Physics Education and Research Center for Dielectric and Advanced Matter Physics \\
Pusan National University, Busan 609-735, Korea}


\begin{abstract}
We review Floquet formalism of quantum electron pumps. In the Floquet formalism the quantum pump 
is regarded as a time dependent scattering system, which allows us to go beyond the adiabatic limit.
It can be shown that the well-known adiabatic formula given by Brouwer can be derived from the adiabatic
limit of Floquet formalism. We compare various physical properties of the quantum pump both in the adiabatic 
and in the non-adiabatic regime using the Floquet theory.
\end{abstract}

\pacs{73.23.-b, 72.10.Bg, 73.50.Pz}

\maketitle


\section{Introduction}

A quantum pump is a device that generates a dc current at zero bias potential through
cyclic change of system parameters. The most direct way to create a dc current was originally 
proposed by Thouless \cite{Thouless83}, who considered a system subjected to a traveling 
wave potential. This can be realized for example with the help of surface acoustic waves 
\cite{Talyanskii97}. Another possibility is to utilize quantum dots. In closed systems operating 
in the Coulomb blockade regime, integer number of electric charge can be transferred by sequential 
changes of barriers like a turnstile \cite{Kouwenhoven91}, whereas in open systems the electron 
pumping can be driven by {\em adiabatic} shape change in the confining potential or other parameters
which affect the interference pattern of the coherent electrons in the device. In contrast to the more
familiar dissipative rectification of ac current, the charge transferred in each cycle of adiabatic
pumping is independent of the period $T$. After a cycle of the adiabatic shape change we return to 
the initial configuration, but the wavefunction may have its phase changed from the initial wavefunction. 
This is the so-called geometric or Berry's phase \cite{Berry84}. This additional phase is equivalent to
some charges that pass through the quantum dot, namely, pumped charge \cite{Altshuler99}.

A basic idea of adiabatic quantum pump was theoretically proposed by Spivak et al. \cite{Spivak95}
in terms of {\em photovoltaic effect}. It was found that the current linearly proportional to the frequency 
exists when voltages on the gates $U_1(t)$ and $U_2(t)$ are arbitrary periodical functions with the same 
period and different initial phases, and the area enclosed by the path in configuration space 
$\{U_1(t),U_2(t)\}$ is not zero. Zhou et al. \cite{Zhou99} demonstrated that there exists adiabatic quantum 
pumping different from the usual photovoltaic effect. Both the amplitude and the sign of the charge transferred
through a sample per period are random sample spcecific (i.e. not quantized) determined by the quantum 
interference. Switkes et al. experimentally
have realized such an adiabatic quantum pumping through an open quantum dot under the shape deformation 
controlled by two ac gate voltages \cite{Switkes99}. At the same time Brouwer \cite{Brouwer98} proposed 
simple and intuitive approach to deal with the quantum pump using adiabatic scattering matrix, which can 
explain many aspects of the experimental results perfomed by Switkes et al.. Since then, adiabatic 
quantum pumps have attracted considerable interest from many researchers \cite{Lubkin99}.

In other sense, the quantum pump is a {\em time dependent} system driven by (at least) two different time 
periodic perturbations with the same angular frequency and initial phase difference $\phi$. One can deal with 
this problem using not only adiabatic approximation exploited by Zhou et al. and by Brouwer but also the 
so-called Floquet approach \cite{Sambe73}. Floquet method allows us to solve the time dependent (mostly periodic)
problem precisely. In physical ponit of view, an oscillating potential can transfer an incoming electron of energy 
$E$ to Floquet side bands at $E\pm n\hbar \omega$, where $n$ is an integer and $\omega$ is the angular 
frequency of the oscillation. A scattering matrix for a time dependent system can be constructed from the 
interplay of these sidebands \cite{Li99,Henseler01}. The Floquet approach can bring us beyond adiabatic
regime of quantum pumps. In this review we will present a breif review on the floquet formalism of quantum 
pump developed recently.


\section{Floquet Formalism for Transport Problem}

Moskalets and B\"uttiker \cite{Moskalets02b}, and Kim \cite{Kim02} have developed the Floquet scattering theory of 
quantum pump independently. First, we breifly introduce the derivation of pumped current under the Floquet theory.

\subsection{Directed charge currents}

Consider the one-dimensional time-dependent Schr\"odinger equation $i\hbar(\partial/\partial t)\psi = H(t)\psi$ 
for a non-interacting electron with mass $\mu$ and $H(t) = -\hbar^2\nabla^2/2\mu + U(x,t)$, where $U(x,t+T)=U(x,t)$ 
and $U(x,t)=0$ at $x\rightarrow \pm\infty$. Due to the periodicity in time, a solution can be written as
\begin{equation}
\Psi_\epsilon(x,t) = e^{-i \epsilon t/\hbar}\sum_{n=-\infty}^{\infty} \chi_n (x) e^{-in\omega t},
\label{floquet}
\end{equation}
where $\epsilon$ is the Floquet energy which takes a continuous value in the interval
$[0,\hbar \omega)$. Since the potential is zero at $x\rightarrow \pm \infty$, 
$\chi_n(x)$ is given by the following form
\begin{equation}
\chi_n (x) = \left\{
\begin{array}{c}
A_n e^{ik_n x} + B_n e^{-ik_n x}, ~~~ x\rightarrow -\infty \\
C_n e^{ik_n x} + D_n e^{-ik_n x}, ~~~ x\rightarrow +\infty,
\end{array} \right.
\label{plane_wave}
\end{equation}
where $k_n=\sqrt{2\mu(\epsilon + n\hbar\omega)}/\hbar$. By wave matching, the incoming and the 
outgoing waves can be connected by matrix $M$
\begin{equation}
\left(
        \begin{array}{c}
        \vec{B} \\ \vec{C}
        \end{array}
\right)
=M
\left(
        \begin{array}{c}
        \vec{A} \\ \vec{D}
        \end{array}
\right).
\label{mat_eq}
\end{equation}
If we keep only the propagating modes ($k_n$ is real), we can obtain the unitary Floquet 
scattering matrix [see the argument below Eq.~(\ref{mat_eq2_ap})], which can be expressed in the following form 
\cite{Li99,Henseler01,Kim03pole}
\begin{equation}
S(\epsilon) = \left(
        \begin{array}{cccccc}
        {\bf r}_{00} & {\bf r}_{01} & \cdots & {\bf t'}_{00} & {\bf t'}_{01} & \cdots \\
        {\bf r}_{10} & {\bf r}_{11} & \cdots & {\bf t'}_{10} & {\bf t'}_{11} & \cdots \\
        \vdots & \vdots & \ddots & \vdots  & \vdots  & \ddots \\
        {\bf t}_{00} & {\bf t}_{01} & \cdots & {\bf r'}_{00} & {\bf r'}_{01} & \cdots \\
        {\bf t}_{10} & {\bf t}_{11} & \cdots & {\bf r'}_{10} & {\bf r'}_{11} & \cdots \\
        \vdots & \vdots & \ddots & \vdots  & \vdots  & \ddots \\
        \end{array}
\right),
\label{osc_smatrix}
\end{equation}
where ${\bf r}_{\alpha\beta}$ and ${\bf t}_{\alpha\beta}$ are the reflection and the 
transmission amplitudes respectively, for modes incident from the left with an Floquet energy 
$\epsilon$ which take continuous values in the interval $[0,\hbar \omega)$; 
${\bf r'}_{\alpha\beta}$ and ${\bf t'}_{\alpha\beta}$ are similar quantities for modes 
incident from the right. 

To get a current let us consider the scattering states. For an energy $E=\hbar^2 k^2/2\mu$ ($k>0$) 
of the incoming particle the scattering states as a solution of the Schr\"odinger equation 
can be defined as
\begin{eqnarray}
\chi^+_E(x,t) &=& 
 	\left\{ 
		\begin{array}{ll}
			e^{ikx-iEt/\hbar} + \sum_{E_n>0} r^+_{E_nE}
				e^{-ik_nx-iE_nt/\hbar}, & x \rightarrow -\infty, \\
			\sum_{E_n>0} t^+_{E_nE} e^{ik_nx-iE_nt/\hbar}, 
				& x \rightarrow +\infty, 
		\end{array}
	\right.
\label{scatt1}
\\
\chi^-_E(x,t) & =& 
	\left\{
		\begin{array}{ll}
			e^{-ikx-iEt/\hbar} + \sum_{E_n>0} r^-_{E_nE}
				e^{ik_nx-iE_nt/\hbar}, & x \rightarrow +\infty, \\
			\sum_{E_n>0} t^-_{E_nE} e^{-ik_nx-iE_nt/\hbar}, 
				& x \rightarrow -\infty, 
		\end{array}
	\right.
\label{scatt2}
\end{eqnarray}
where $E_n = E+n\hbar\omega$, $k_n=\sqrt{2\mu E_n}/\hbar$, and the normalization is ignored. 
Here we have reflection and transmission coefficients $r^+_{E_nE}$ and $t^+_{E_nE}$, 
which can be obtained from the {\em unitary} Floquet scattering matrices $S$ obtained above. 
\cite{Kim02,Henseler01,Kim03pole}. The above transmission and reflection coefficients are related to 
the matrix elements of $S$ in terms of 
$t^+_{E_n,E} = \sqrt{k_\beta/k_\alpha}{\bf t}_{\alpha\beta}$, {\it etc.} with 
$E=\epsilon+\beta\hbar\omega$ and $\alpha=n+\beta$.

We derive the current using these scattering states. The time dependent electron field 
operator can be obtained in the following form \cite{Buettiker92,Levinson99,Kim03pauli}
\begin{equation}
\Psi(x,t) = \sum_\sigma \int dE ~ \chi^\sigma_E(x,t) \frac{a_{\sigma E}}{\sqrt{hv(E)}},
\label{expansion}
\end{equation}
where $a_{\sigma E}$ and $v(E)$ is an annihilation operator for electrons in the scattering 
states $\chi^\sigma_E(x,t)$ and the velocity, respectively. Even though the scattering states 
do not form othogonal bases we need only the completeness to be sure that the expansion of 
Eq.~(\ref{expansion}) is valid. Using this field operator the current operator 
is also expressed as
\begin{eqnarray}
J(x,t) & = & (ie/2m)\Psi^+(x,t)\nabla \Psi(x,t) + H.c. \nonumber \\
 &= &\frac{e}{m}\sum_{\sigma\sigma'} \int dE dE'
       ~ {\rm Im}(\chi^{\sigma' *}_{E'} \nabla \chi^{\sigma}_E)
        \frac{a^+_{\sigma E} a_{\sigma' E}} {h\sqrt{v(E)v(E')}}.
\end{eqnarray}

The quantum mechanical (or thermal) average of the current operator becomes
\begin{equation}
\left< J(x,t) \right> 
 = \frac{e}{m}\sum_\sigma \int dE ~ {\rm Im}(\chi^{\sigma *}_{E} \nabla \chi^{\sigma}_E) 
   \frac{\left< a_{\sigma E}^+a_{\sigma E} \right>}{hv(E)} 
 + \frac{e}{m}\sum_{\sigma \sigma'} \sum_{E_n>0, n \neq 0} \int dE \label{current_avg}
  {\rm Im}(\chi^{\sigma' *}_{E_n} \nabla \chi^{\sigma}_E)
      \frac{\left< a_{\sigma' E_n}^+a_{\sigma E} \right>}{h\sqrt{v(E)v(E_n)}}. 
\end{equation}
We evaluate Eq.~(\ref{current_avg}) taking $x \rightarrow \infty$ and averaging over 
space and time. One can then obtain the pumped current as following
\begin{equation}
I = \frac{e}{h}\sum_{E_n>0}\int dE \left[T^+_{E_nE}f_L(E) - T^-_{E_nE}f_R(E)\right],
\label{final}
\end{equation}
where we exploit $\sum_{E_n} (k_n/k)|r^-_{E_nE}|^2= 1-\sum_{E_n} (k_n/k)|t^-_{E_nE}|^2$ and 
$\left< a_{\sigma E}^+a_{\sigma E} \right> = f_\sigma(E)$. The second term of the righthand 
side in Eq.~(\ref{current_avg}) vanishes due to the unitarity of the scattering matrix
\cite{cal_detail}. Here, $T^{\pm}_{E_nE}$ denotes $(k_n/k)|t^{\pm}_{E_nE}|^2$.
The finally obtained current expression (\ref{final}) simply means that the pumped current corresponds
to the difference beween the currents going from the right to the left and from the left to the right.

\subsection{Adiabatic limit}

Moskalets and B\"uttiker \cite{Moskalets02b} have shown that the adiabatic limit of the Floquet formalism 
is equivalent to the adiabatic scattering matrix approach given by Brouwer. Here we summarize it.

The adiabatic condition in the quantum pump implies that any time scale of the problem 
considered must be much smaller than the period of the oscillation of an external pumping 
\cite{Brouwer98}. We can then define the instantaneous scattering matrix with time dependent 
parameters, namely $X_n(t)$,
\begin{equation}
\hat{S}_{ad}(E,t) = \hat{S}_{ad}(E,\{X_n(t)\})=
\left(
        \begin{array}{cc}
        \hat{r}_{ad} & \hat{t}'_{ad} \\ 
        \hat{t}_{ad} & \hat{r}'_{ad} 
        \end{array}
\right)
\label{adiabatic}
\end{equation}
Due to the time periodicity of $X_n$'s, using a Fourier transform one can obtain the amplitudes
of side bands for particles traversing the adiabatically oscillating scatterer with incident
energy $E$ as following
\begin{equation}
\hat{S}_{ad}(E,\{X_n(t)\}) = \sum_n \hat{S}_{ad,n}(E)e^{-in\omega t},
\label{adiabatic1}
\end{equation}
where
\begin{equation}
\hat{S}_{ad,n}(E)=\frac{1}{T}\int_0^T dt~e^{in\omega t} \hat{S}_{ad}(E,\{X_n(t)\}).
\label{adiabatic2}
\end{equation}
Thus we can construct the adiabatic Floquet scattering matrix as follows 
\begin{equation}
S(E_n,E) \approx S(E,E_{-n}) \equiv \hat{S}_{ad,n}(E),
\label{moska_adia}
\end{equation}
where $E_n=E+n\hbar\omega$. 

Now we go back to the current expression Eq.~(\ref{final}) in the Floquet formalism. Assuming
$f_L(E)=f_R(E) \equiv f(E)$, Eq.~(\ref{final}) can be written as
\begin{equation}
I=\frac{e}{h}\sum_{E_n>0}\int dE 
\left[ |S^{+}_{E_nE}|^2 f(E) - |S^{-}_{EE_{-n}}|^2 f(E_{-n}) \right],
\label{adia1}
\end{equation}
where we make the shift $E \rightarrow E-n\hbar\omega$ for the second term in the bracket.
Here one can see $T^\pm_{E'E}=|S^\pm_{E'E}|^2 $ 
since ${\bf t}_{nm} ({\bf t'}_{nm})= S^+_{E'E} (S^-_{E'E})$ with
$E'=\epsilon+n\hbar\omega$ and $E=\epsilon+m\hbar\omega$.
Using the adibatic approximation Eq.~(\ref{moska_adia}) and $\omega << 1$, we obtain
 \begin{equation}
I_{ad}=\frac{\omega e}{2\pi} \int dE \frac{\partial f(E)}{\partial E}
\sum_{E_n>0} n |\hat{S}_{ad,n}(E)|^2.
\label{adia2}
\end{equation}
Using Eq.~(\ref{adiabatic1}), Eq.~(\ref{adia2}) can be rewritten as follows:
\begin{equation}
I_{ad}=i\frac{\omega e}{4\pi^2} \int_0^T dt \int_0^\infty dE \frac{\partial f(E)}{\partial E}
\rm{Tr} \left( \frac{\partial \hat{S}_{ad}(E,t)} {\partial t} \hat{S}^\dagger_{ad}(E,t) \right).
\end{equation}
This is Brouwer's formula.

\subsection{Quantum pumping and broken symmetry}

In the adiabatic limit the pumped current occurs when the area enclosed by the path in the parameter space
$\{ V_1(t), V_2(t) \}$ is not equal to zero. It implies that the system breaks the time reversal symmetry. The
time reversability of the time dependent system is subtle. We define that the system is time reversible if the
Hamiltonian satisfies $H(t) = H(-t)$ after appropriate time translation $H(t+\alpha)$ with a certain $\alpha$.
For simplicity let us assume the potential $V({\bf r},t)$ is given by
\begin{equation}
V({\bf r},t,\phi) = V_1({\bf r}) \cos(\omega t-\phi/2) + V_2({\bf r}) \cos(\omega t +\phi/2), 
\label{potential_cos}
\end{equation}
where $V_1$ and $V_2$ represent the spatial dependence of two time dependent perturbations. $\phi$ is the 
initial phase difference of two time-dependent perturbations. In case $\phi=0$ one finds that 
\begin{equation}
V({\bf r},t,\phi) = \left[ V_1({\bf r}) + V_2({\bf r}) \right] \cos \omega t,
\label{potential_cos1}
\end{equation}
which means effectively there is only one time dependent perturbation, so that no pumped current exists.
Even though the cosine function is replaced by sine, by adding $\pi/2$ to the time $t$, one can make the
system retain time reversal symmetry. It is emphasized that in the adiabatic limit the time reversal symmetry
should be broken in oder to obtain the pumped current.

In the non-adiabatic case, however, the condition that the pumped current appears is quite different from
that in the adiabatic limit. From Eq.~(\ref{final}) one can see
the pumped current exists when $T^+_{E_n,E} \neq T^-_{E_n,E}$
is fulfilled. We assume that the system has the time reversal symmetry. The time reversed process of the
transition $T^+_{E_n,E}$ is given by $T^-_{E,E_n}$ since not only one should change the directions of the 
momenta but also replace the emission by the absorption, and vice versa. It means that for the system with the 
time reversal symmetry $T^+_{E_n,E} = T^-_{E,E_n}$ holds. This is nothing but $S^T = S$, where $T$ represents
the transpose of a matrix. However, such an equality does not guarantee $T^+_{E_n,E} = T^-_{E_n,E}$. 
In other words, the pumped current in the non-adiabatic case can still exist even when the time reversal 
symmetry holds. In the adiabatic limit (namely, $\omega \rightarrow 0$), $T^+_{E_n,E}$ approaches 
$T^+_{E,E_n}$. Then the time reversal symmetry, i.e. $T^+_{E,E_n} = T^-_{E_n,E}$, leads us to 
$T^+_{E_n,E} \approx T^-_{E_n,E}$, i.e. no pumped current. 
This is the reason why the pumped current vanishes in the adiabatic 
limit when the time reversal symmetry is preserved.

\begin{figure}
\center
\includegraphics[height=8 cm,angle=0]{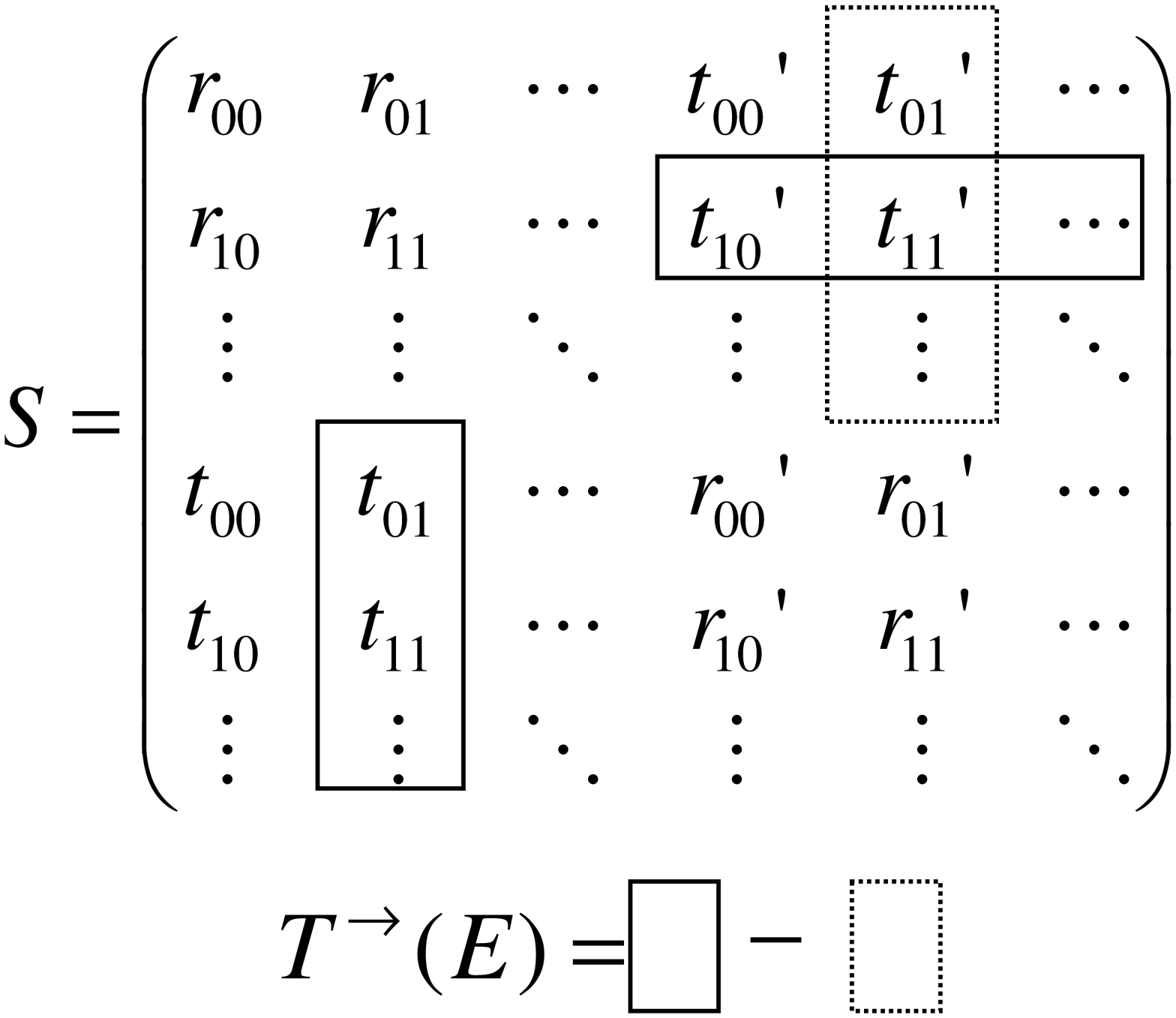}
\caption{Even though the system has time reversal symmetry ($S=S^T$), the transmission probabilities for
the opposite directions can differ from each other. Note that in Floquet formalism the transmission probability is 
obtained from taking summation of {\em one column}.}
\label{fig1}
\end{figure}

To get more physical intuition for the relation between the pumped current and the broken symmetry let us 
reconsider the current expression given in Eq.~(\ref{final}). The pumped current $I$ is given by the difference
of two currents having the opposite directions. The transmission probability to the right for an electron with the 
incoming energy $E$ is $\sum_{E_n>0} T^+_{E_nE}$ ($\equiv T^+_E$). Comparing it with Eq.~(\ref{osc_smatrix}) 
one can find that $T^+_E$ correponds to the summation of the $m$th column of a part of the Floquet scattering 
matrix when $E=\epsilon+m\hbar\omega$ ($m$ is an integer). More precisely 
$\sum_{E_n>0} T^+_{E_nE}=\sum^\infty_{n=0} {\bf t}_{nm}$. It is worth mentioning that in usual multi-channel
scattering problems with {\em time-independent} potentials the transmission probability is given by
the summation of a block of a scattering matrix, e.g. $\sum_{mn} {\bf t}_{nm}$, where $m$ and $n$
represents the number of modes or scattering channels of the incoming and the outgoing lead, respectively.
Even though the system considered has time reversal symmetry, i.e. $S=S^T$ or equivalently 
${\bf t}_{nm} = {\bf t'}_{mn}$, it is not guaranteed that $\sum_n {\bf t}_{nm} = \sum_n {\bf t'}_{nm}$
(see Fig~.1), while it is always true that $\sum_{nm} {\bf t}_{nm} = \sum_{nm} {\bf t'}_{nm}$ once 
${\bf t}_{nm} = {\bf t'}_{mn}$ is fulfilled, which is related to the reciprocity. The physical meaning of the 
summation of one column in the case with the oscillating potential to obtain $T^\pm_E$ is the following. 
The incoming electron has a definite incident energy $E$ since the energy is well defined far 
from the scatterer, whereas the outgoing electron can have various energies to be taken into account since 
an oscillating potential can transfer the electron with energy $E$ to side bands at $E \pm \hbar\omega$.

\subsection{Wigner delay time}

The adiabatic approximation implies that any time scale of the problem considered, especially the electron dwell 
time in a quantum pump must be much smaller than the period of the oscillation of a external pumping $T_p$
\cite{Brouwer98}. Using the Floquet formalism we can calculate the Wigner delay time $\tau_W$, 
which is the interaction time of the incident electron with the scattering potential \cite{Wigner55,Smith60,Fyodorov97}. 
In this sense $\tau_W$ corresponds to the electron dwell time in the quantum pump. To obtain the Wigner delay time 
we use the eigenvalues of the Floquet scattering matrix $S$. Due to the unitarity of $S$ all the eigenvalues lie on the 
unit circle and can be written in the form $\exp(i\theta_\alpha)$. The Wigner delay time is defined by
\begin{equation}
\tau_W = \hbar \sum_\alpha \frac{d\theta_\alpha}{dE} \left|\left< k_n |\theta_{\alpha} \right>\right|^2,
\label{wigner_eq}
\end{equation}
where the eigenstate corresponding to the eigenvalue $\theta_{\alpha}$ and an input propagating 
state (or channel) with momentum $k_n$ are denoted by $\left| \theta_{\alpha} \right>$ and 
$\left| k_n \right>$ respectively \cite{Emmanouilidou01}. It is worth noting that the Wigner 
delay time is a function of the energy of the incident particle $E$ ($=E_{Fl}+n\hbar\omega$),
and $\left| k_n \right>$ and $\left| \theta_{\alpha} \right>$ are determined by $n$ and
$E_{Fl}$ respectively. If $\left< k_n |\theta_{\alpha} \right>$ is ignored in Eq.~(\ref{wigner_eq}) 
the Wigner delay time $\tau_W$ becomes trivial, i.e. $\tau_W(E+n\hbar\omega) = \tau_W(E)$. 


\section{Pauli Blocking Factors}

In mesoscopic systems it is well known that the current through the scatterer can be obtained from
\begin{equation}
I = \frac{e}{h}\int dE dE' \left[ T^+(E',E)f_L(E) - T^-(E,E')f_R(E') \right],
\label{no pauli current}
\end{equation}
where $T^+(E',E)$ represents the transmission probability for scattering states incident 
from the left at energy $E$ and emerging to the right at $E'$, and $T^-(E,E')$ is defined 
in a similar manner for the reverse direction. $f_L$ ($f_R$) is the Fermi-Dirac 
distribution in the left (right) reservoir. There has been some debate on using this 
formula \cite{Datta95,Hekking91,Landauer92,Datta92a,Datta92b,Sols92,Boenig93,Wagner00}
since Eq.~(\ref{no pauli current}) dose not contain the so-called Pauli blocking factors.
The fermionic nature of the electrons is taken care of in an {\it ad hoc} way by factors
$1-f$ to suppress scattering into occupied states, so that the current is given by
\begin{equation}
I = \frac{e}{h}\int dE dE' \left\{ T^+(E',E)f_L(E) \left[ 1-f_R(E') \right] \right. 
  - \left. T^-(E,E')f_R(E') \left[ 1-f_L(E) \right] \right\}.
\label{pauli current}
\end{equation}
Usually these two expressions Eqs.~ (\ref{no pauli current}) and (\ref{pauli current}) give 
the same results since the difference between them,
$[T^+(E',E)-T^-(E,E')]f^L(E)f^R(E')$, vanishes when $T^+(E',E)=T^-(E,E')$, i.e., the 
micro-reversibility holds. The question can arise, however, if the system lack of this
micro-reversal symmetry is considered. One of the relevant example is a quantum pump.

More than two periodically oscillating perturbations with a phase difference
($\neq n\pi$, $n$ is an integer) break the time reversal symmetry \cite{Wagner00}
as discussed in Sec.II.C, and 
consequently $T^+(E',E) \neq T^-(E,E')$. Therefore, the currents obtained from 
Eq.~(\ref{no pauli current}) and (\ref{pauli current}) are different from each other 
in quantum pumps \cite{Datta92b,Wagner00}. The question immediately arises: which one 
is correct in quantum pumps? It is noted that this problem still exists even in the 
adiabatic limit. We would like to make a conclusion first. 
Even though the time dependent scatterer like the case of quantum pumping not only 
cause inelastic scatterings but also break the mico-reversibility, 
{\em the Pauli blocking factor is unnecessary} when the scattering process through 
the scatterer is {\em coherent}. 

The existence of the Pauli blocking factors is intimately related to the ``scattering
states'' \cite{Datta95}. If we fill up the energy eigenstates with the electrons in both 
reservoirs independently and then transfer the electrons  from one to the other reservoir, 
the Pauli blocking factors cannot be avoided. If the transport is coherent across the 
scatterer, however, one can define a single wavefunction extending from one reservoir to the 
other (more precisely reflected and transmitted waves in every connected reservoirs) and 
then fill up these scattering states. In this consideration the concept of transferring 
the electron from one to the other reservoir is automatically eliminated, so is the Pauli 
blocking. In Sec.~II.A using the scattering states given
in Eq.~(\ref{scatt1}) and (\ref{scatt2}) we have already shown that the current 
experssion (\ref{final}) does not contain Pauli blocking factor.

Recently, Wagner has raised the question on the Pauli blocking factor in quantum pumps,
and concluded that a marked difference is predicted for the temperature dependence of
the pumped current for the case with and without the Pauli blocking factor \cite{Wagner00}. 
The pumped current can be generated from two contributions: classical and quantum origin.
Most interest has been focused on the quantum mechanical charge pumping in chaotic quantum
dots, where the pumped current has zero average over dot configurations and shows the 
mesoscopic fluctuation \cite{Brouwer98,Switkes99}. The important measure quantifying 
the currents pumped quantum mechanically is a mean square average $\left<I^2\right>^{1/2}$. 
Since the mesoscopic fluctuation of the pumped current results from the electron interfence 
and is generically quantum nature, the scattering process in physically meaningful quantum 
pumps should be coherent. It was observed in experiment \cite{Switkes99} that as the 
temperature increases the mean square average decreases due to the temperature-dependent
decoherence. The Pauli blocking factor is then unnecessary from the beginning, which is 
also consistent with the results obtained from non-equilibrium Green's function formalism 
in Ref. \cite{Datta92b}.


\section{Magnetic Field Inversion Symmetry}

There has been the puzzle unsolved in the experiment performed by Switkes et al. \cite{Switkes99}, 
which is the magnetic field inversion symmetry (MIS) in pumped currents. It was suggested theoretically 
\cite{Zhou99} that the pumped current $I$ is invariant upon magnetic field reversal
\begin{equation}
I(B) = I(-B).
\label{symmetry}
\end{equation}
The subsequent experiment on open quantum dots appears to be in a good agreement with 
Eq.~(\ref{symmetry}) \cite{Switkes99}. It became immediately clear, however, that such 
symmetry is not valid in theory \cite{Shutenko00}. In addition, the measured dc voltage is 20 times 
larger than the theoretical estimation. Furthermoere, the voltage fluctuation decreases upon applying 
the static magnetic field breaking time-reversal symmetry, which seems to be the weak localization-like 
phenomenon of disordered systems, while such an effect cannot be predicted in the theory. Brouwer 
proposed that the rectified displacement current could account for all these discrepancies  \cite{Brouwer01}, 
which implies the dc current observed in the experiment is not attributed to the adiabatic quantum pumping. 
Recently Marcus group has reported that the origin of the dc current is mesoscopic rectification as well as 
adiabtic quantum pumping \cite{DiCarlo03}. Strictly speaking, a reliable adiabatic quantum pumping has 
not been realized yet.

\begin{figure}
\center
\includegraphics[height=12 cm,angle=0]{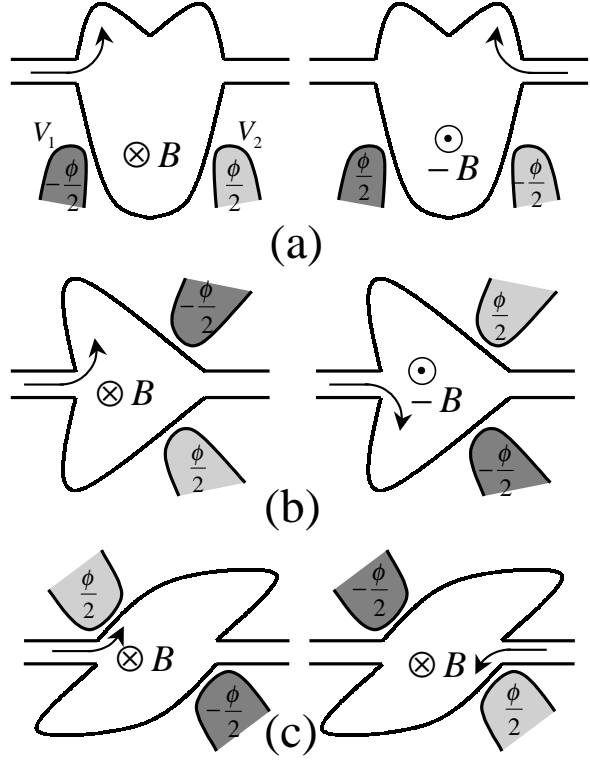}
\caption{The schematic diagram for three discrete symmetries of quantum pumps, i.e. (a) LR, 
(b) UD and (c) IV symmetry.}
\label{fig2}
\end{figure}

Even though in general quantum pumps do not fulfill the MIS, additional discrete symmetries 
of quantum dots can lead to such MIS. The effect of discrete symmetries on the MIS has been 
studied by Aleiner, Altshuler and Kamenev \cite{Aleiner00} in the adiabatic limit. They found 
the reflection symmetries give rise to relations $I(B)=I(-B)$ or $I(B)=-I(-B)$ depending on 
the orientation of the reflection axis. In the presense of inversion $I(B)=0$. Note that the 
symmetry considered in Ref. \cite{Aleiner00} should be kept intact in the pumping cycle,
and the results obtained are available only in the {\em adiabatic} regime.
Using the Floquet approach we can also investigate the MIS of the pumped 
current for three discrete symmetries, namely LR (left-right), UD (up-down), and IV (inversion), 
as shown in Fig.~\ref{fig2}, {\em both in the adiabatic and in the non-adiabatic cases}. 
These symmetries are only applied to the {\em time independent} part of the scattering 
potential, which is $V_n({\bf r})$ in Eq.~(\ref{potential}), with an arbitrary $\phi$, so that 
they are not necessarily kept intact during the pumping cycle \cite{Kim03mag}.

\subsection{Floquet approach}

Let us consider the two-dimensional quantum dot driven by time-periodic perturbations 
under a static magnetic field, which is attached to two leads (see Fig.~\ref{fig2}), 
whose Hamiltonian is given by
\begin{equation}
i\hbar \frac{\partial\psi}{\partial t} 
= \left[\frac{(i\hbar\nabla + eA)^2}{2\mu} +V({\bf r},t)\right]\psi,
\label{schro_b}
\end{equation}
where $A$ is the vector potential. The potential $V({\bf r},t)$ is given by 
\begin{equation}
V({\bf r},t,\phi) = V_0({\bf r}) 
+ V_1({\bf r}) \cos(\omega t-\phi/2) + V_2({\bf r}) \cos(\omega t +\phi/2), 
\label{potential}
\end{equation}
where $V_0$ represents the confined potential of a quantum dot, and $V_1$ and $V_2$ are
the spatial dependence of two time dependent perturbation. $\phi$ is the initial phase 
difference of two time-dependent perturbations, and ${\bf r}$ denotes $(x,y)$. Note that 
$V({\bf r},-t,\phi)=V({\bf r},t,-\phi)$. When $\phi=n\pi$ without magnetic field, the 
quantum pump is microscopically reversible (or time reversible), i.e. 
$t^{\beta\alpha}_{nm} = t'^{\alpha\beta}_{mn}$, where $t_{nm}^{\beta\alpha}$ represents 
the transmission amplitude from the Floquet side band $m$ of the channel $\alpha$ in the 
left lead to the side band $n$ of the channel $\beta$ in the right lead, and 
$t'^{\alpha\beta}_{mn}$ is a similar quantity for the opposite direction. In a quantum
pump the two time dependent perturbations with a finite $\phi$ ($\neq n\pi$) as well 
as the magnetic field break the time reversal symmetry as already discussed in Sec.~II.C

If we take the complex conjugate of Eq.~(\ref{schro_b}) and at the same time, reverse 
the vector potential ($A \rightarrow -A$) and time ($t \rightarrow -t$), we obtain
\begin{equation}
i\hbar \frac{\partial\psi^*}{\partial t} 
= \left[\frac{(i\hbar\nabla + eA)^2}{2\mu} +V({\bf r},-t)\right]\psi^*.
\end{equation}
Compared with Eq.~(\ref{schro_b}) one can see that 
$\psi_{-B,\phi}({\bf r},t) =\psi^*_{B,-\phi}({\bf r},-t)$. 
The complex conjugation combined with the time inversion of the solution in Eqs.~(\ref{floquet}) 
and (\ref{plane_wave}) simply corresponds to the reversal of the momentum direction of the 
incoming or the outgoing plane waves and simultaneously the replacement of absorption by 
emission, and vice versa. This simply implies that $S \rightarrow S^T$ ($S^T$ is the transpose 
of $S$), i.e. $S_{nm}^{\beta\alpha} \rightarrow S_{mn}^{\alpha\beta}$. 
For completion of the magnetic field inversion operation, $\phi \rightarrow -\phi$ also has to 
be considered.

\begin{figure}
\center
\includegraphics[height=6.5 cm,angle=0]{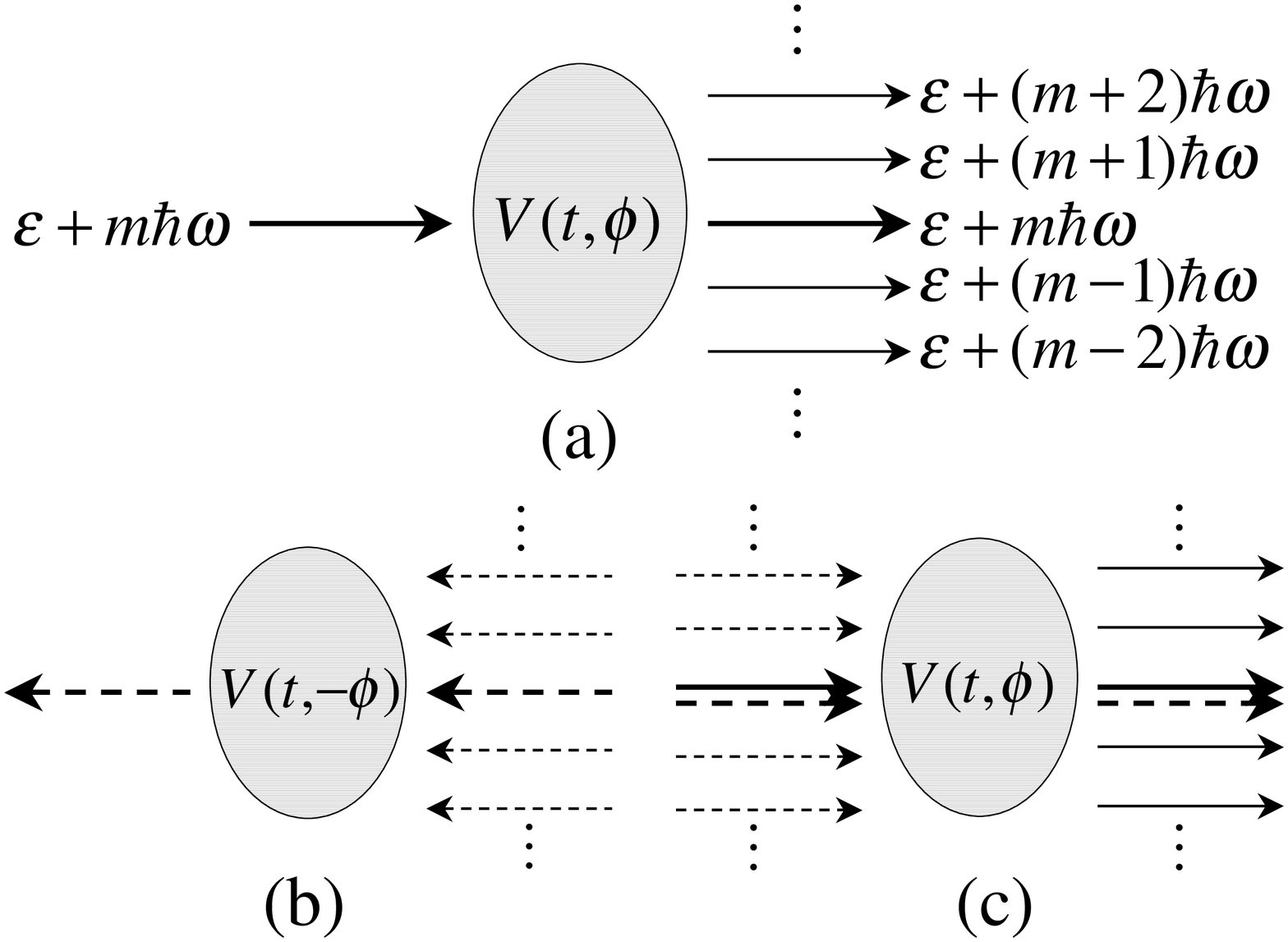}
\caption{Schematic diagrams for describing the configuration of side bands with given channels 
$\alpha$ and $\beta$ to calculate the total transmission coefficients through the time-periodic 
perturbation $V(t)$ for (a) $T^\rightarrow_{B,\phi} (E)$ and (b) the case given in 
Eq.~(\ref{eq-b}). (c) The setup (a) (the solid arrows) superimposed by the setup (b) reversed 
by using IV symmetry given in Eq.~(\ref{IVsym_coef}) (the dashed arrows).}
\label{fig3}
\end{figure}

The total transmission coefficient to the right reservoir after the magnetic field inversion is given by
\begin{eqnarray}
T^+_{-B,\phi}(E) & = & 
\sum_{\alpha\beta}\sum_{n=0}^{\infty}\left| t_{nm}^{\beta\alpha}(\epsilon,-B,\phi)\right|^2 
\label{eq-a}\\
& = & \sum_{\alpha\beta}\sum_{n=0}^{\infty}
\left| t'^{\alpha\beta}_{mn}(\epsilon,B,-\phi)\right|^2, \label{eq-b}
\end{eqnarray}
where $E=\epsilon+m\hbar\omega$. It must be noted that Eq.~(\ref{eq-b}) is not equivalent to 
$T^-_{B,-\phi}$. Figures \ref{fig3}(a) and (b) represent the usual setup of side 
bands for calculating the total transmission coefficients, e.g. $T^+_{B,\phi}(E)$, 
and the setup for calculating Eq.~(\ref{eq-b}), respectively. In general, they are different 
from each other, so are $I(B)$ and $I(-B)$. The MIS is not valid in Floquet formalism, as
well.

{\bf LR symmetry}.
If we assume LR symmetry [$V_0(x,y)=V_0(-x,y)$ and $V_1(x,y)=V_2(-x,y)$], the following 
relations are obtained [see Fig.~\ref{fig2}(a)] 
\begin{eqnarray}
t^{\beta\alpha}_{nm}(\epsilon,B,\phi) & = & t'^{\beta\alpha}_{nm}(\epsilon,-B,-\phi), \nonumber \\
r^{\beta\alpha}_{nm}(\epsilon,B,\phi) & = & r'^{\beta\alpha}_{nm}(\epsilon,-B,-\phi).
\label{LRsym_coef}
\end{eqnarray}
These are expected from the symmetries of classical trajectories \cite{Linke00}.
Eq.~(\ref{eq-a}) can then be rewritten as
\begin{eqnarray}
T^+_{-B,\phi}(E) & = & \sum_{\alpha\beta}\sum_{n=0}^{\infty}
\left| t'^{\beta\alpha}_{nm}(\epsilon,B,-\phi)\right|^2 \nonumber \\
& = & T^-_{B,-\phi}(E).
\label{LRsym_final}
\end{eqnarray}
From Eq.~(\ref{final}) one reaches $I(B,\phi)=-I(-B,-\phi)$. 
Since such relation results from the Floquet approach, it is available for both the adiabatic 
and the non-adiabatic cases. At zero magnetic field one finds $I(\phi)=-I(-\phi)$, consequently 
$I(\phi=0)=0$ even for the non-adiabatic case.

{\bf UD symmetry}.
For UD symmetry [$V_0(x,y)=V_0(x,-y)$ and $V_1(x,y)=V_2(x,-y)$], from the similar procedure used in 
Eq.~(\ref{LRsym_coef}) and (\ref{LRsym_final}) one finds $I(B,\phi)=I(-B,-\phi)$. 
Like the LR symmetry, such symmetry of the quantum pump is correct for both the adiabatic and the 
non-adiabatic cases. At zero magnetic field one finds $I(\phi)=I(-\phi)$, which can be understood from 
the fact that the exchange of $V_1$ and $V_2$, i.e. $\phi \rightarrow -\phi$, does not make any difference 
in Fig.~\ref{fig2}(b).

{\bf IV symmetry}.
If one assume IV symmetry [$V_0(x,y)=V_0(-x,-y)$ and $V_1(x,y)=V_2(-x,-y)$], one obtains 
[see Fig.~\ref{fig2}(c)]
\begin{eqnarray}
t^{\beta\alpha}_{nm}(\epsilon,B,\phi) = t'^{\beta\alpha}_{nm}(\epsilon,B,-\phi), \nonumber \\
r^{\beta\alpha}_{nm}(\epsilon,B,\phi) = r'^{\beta\alpha}_{nm}(\epsilon,B,-\phi).
\label{IVsym_coef}
\end{eqnarray}
Eq.~(\ref{eq-b}) is reexpressed as
\begin{equation}
T^+_{-B,\phi}(E) = \sum_{\alpha\beta}\sum_{n=0}^{\infty}
\left| t_{mn}^{\alpha\beta}(\epsilon,B,\phi)\right|^2,
\end{equation}
which is different from $T^+_{B,\phi}(E)$ since the summation is taken for $n$
as shown in Fig.~\ref{fig3}(c). In general, there is no relevant MIS for dots with IV symmetry. 
When $\phi=0$, however, Eq.~(\ref{IVsym_coef}) leads to $T^+_B(E)=T^-_B(E)$,
consequently $I=0$. 

\subsection{Adiabatic limit}

We have already investigated the relation between the Floquet and the adiabatic approach in Sec.~II.B.
Once the adiabatic scattering matrix is found, by using Brouwer's formula \cite{Brouwer98} 
the pumped charge $Q$ per one cycle can be expressed as
\begin{equation}
Q(B)=g[\hat{t}_{ad}(B),\hat{r}'_{ad}(B)] \equiv \int^T_0 dt \{f[\hat{t}_{ad}(B)]+f[\hat{r}'_{ad}(B)]\},
\label{Brouwer_eq}
\end{equation}
where
\begin{equation}
f(s)=\frac{e}{2\pi}\sum^2_{k=1}\sum_{\alpha\beta} {\rm Im}
\frac{\partial s_{\alpha\beta}}{\partial X_k}s^*_{\alpha\beta} \frac{dX_k}{dt}.
\end{equation}
Then, the pumped current $I$ is $Q/T$. Below, we will use Eq.~(\ref{Brouwer_eq}) for the
current. We introduce $I_r$ and $I_l$, which are the pumped current to the right and the left, respectively, 
so that $I_r=-I_l$. $I_r$ is nothing but $I$ in the previous notation.

{\bf LR symmetry}.
The relation (\ref{LRsym_coef}) implies $\hat{t}(\hat{r})^{\beta\alpha}_{ad,n}(E;B,\phi)
=\hat{t}'(\hat{r}')^{\beta\alpha}_{ad,n}(E;-B,-\phi)$ in the adiabatic limit, which, by using 
Eq.~(\ref{adiabatic1}), immediately leads to $\hat{t}(\hat{r})_{ad}(E,t;B,\phi)=
\hat{t}'(\hat{r}')_{ad}(E,t;-B,-\phi)$. From Eq.~(\ref{Brouwer_eq}) one finds 
\begin{eqnarray}
I_r(-B,\phi) & = & g[\hat{t}(-B,\phi),\hat{r}'(-B,\phi)] \nonumber \\
& = & g[\hat{t}'(B,-\phi),\hat{r}(B,-\phi)] \\
& = & I_l(B,-\phi) = -I_r(B,-\phi). \nonumber
\end{eqnarray}
Note that if the relation $I(\phi)=-I(-\phi)$, which is available only in the adiabatic limit, is considered, 
one reaches simpler form $I(B)=I(-B)$.

{\bf UD symmetry}.
Like the previous subsection one can obtain
$\hat{t}(\hat{r})^{\beta\alpha}_{ad,n}(E;B,\phi)=\hat{t}(\hat{r})^{\beta\alpha}_{ad,n}(E;-B,-\phi)$ 
in the adiabatic limit, which, by using Eq.~(\ref{adiabatic1}), leads to 
$\hat{t}(\hat{r})_{ad}(E,t;B,\phi)=\hat{t}(\hat{r})_{ad}(E,t;-B,-\phi)$. From Eq.~(\ref{Brouwer_eq}) 
one obtains 
\begin{eqnarray}
I_r(-B,\phi) & = & g[\hat{t}(-B,\phi),\hat{r}'(-B,\phi)] \nonumber \\
& = & g[\hat{t}(B,-\phi),\hat{r}'(B,-\phi)] \\
& = & I_r(B,-\phi). \nonumber
\end{eqnarray}
Note also that if the relation $I(\phi)=-I(-\phi)$ is considered, one reaches simpler form $I(B)=-I(-B)$.

{\bf IV symmetry}.
The relation (\ref{IVsym_coef}) implies $\hat{t}(\hat{r})^{\beta\alpha}_{ad,n}(E;B,\phi)
=\hat{t}'(\hat{r}')^{\beta\alpha}_{ad,n}(E;B,-\phi)$ in the adiabatic limit, which, by using 
Eq.~(\ref{adiabatic1}), leads to $\hat{t}(\hat{r})_{ad}(E,t;B,\phi)=
\hat{t}'(\hat{r}')_{ad}(E,t;B,-\phi)$. Considering the magnetic field inversion operation, 
one can obtain 
\begin{eqnarray}
\hat{t}_{ad}(E,t;B,\phi) & = & \hat{t}^T_{ad}(E,t;-B,\phi), \nonumber \\
\hat{r}_{ad}(E,t;B,\phi) & = & \hat{r}'^T_{ad}(E,t;-B,\phi). 
\label{adia_sym}
\end{eqnarray}
From Eq.~(\ref{Brouwer_eq}) one finds 
\begin{eqnarray}
I_r(-B,\phi) & = & g[\hat{t}(-B,\phi),\hat{r}'(-B,\phi)] \nonumber \\
& = & g[\hat{t}^T(B,\phi),\hat{r}^T(B,\phi)]. 
\end{eqnarray}
There is no relevant symmetry in the adiabatic limit, again. 

Eq.~(\ref{moska_adia}) implies that $t_{m+n,m} \approx t_{m,m-n}$ in the adiabatic limit. 
After summing up all the side band contributions, the two configurations shown in 
Fig.~\ref{fig3}(c) (the solid and the dashed arrows) give the approximately equivalent total 
transmission coefficients, i.e. 
$T^+_{B}(E) \approx T^+_{-B}(E)$. By applying similar procedure to the 
reflection coefficient one can also find $R^+_{B}(E) \approx R^-_{-B}(E)$ 
(make sure that the arrow is now reversed). It is also satisfied that 
$T^+_{B}(E) \approx T^-_{-B}(E)$ since the unitarity of Floquet scattering 
matrix requires $T^{\pm}+R^{\pm}=1$. From these two
relations the pumped current is shown to be vanishingly small, 
$I(B)=\int dE [T^+_B(E) -T^-_B (E)] \approx 0$ since 
$T^+_{B}(E) \approx T^+_{-B}(E) \approx T^-_{B}(E)$.
In Table~I, we summarize the MIS of the pumped currents with various situations for the
three symmetries.

One possibility to realize the experimentally observed MIS, i.e. $I(B)=I(-B)$, is that
the quantum dot would obey LR symmetry. Then, the MIS is recovered only {\em in the
adiabatic limit}, while it will be broken in non-adiabatic case. In general, however, 
it is difficult that a chaotic quantum dot possesses any discrete symmetry like the LR 
symmetry.

\begin{table}
\caption{The symmetries of the pumped currents for the magnetic field inversion.
TRS denotes time reversal symmetry by two time dependent perturbations, i.e.
TRS for $\phi=n\pi$ and no TRS for $\phi \neq n\pi$. $\times$ represents that
there is no relevant symmetry.}
\begin{tabular}{ccccc}
\hline
 & \multicolumn{2}{c}{adiabatic} & \multicolumn{2}{c}{non-adiabatic} \\
 & TRS & no TRS & TRS & no TRS \\ \hline
LR & $I=0$ & $I(B)=I(-B)$ & $I(B)=-I(-B)$ & $I(B,\phi)=-I(-B,-\phi)$ \\
UD & $I=0$ & $I(B)=-I(-B)$ & $I(B)=I(-B)$ &  $I(B,\phi)=I(-B,-\phi)$ \\
IV & $I=0$ & $I\approx 0$ & $I=0$ &  $\times$ \\
\hline
\end{tabular}
\end{table}


\section{Simple example: two oscillating delta-function barriers}

The minimal model that shows quantum pumping effect is one-dimensional (1D) two harmonically 
oscillating delta-functions barriers. Wei et al. have applied Brouwer's adiabatic approach to this model 
to calculate the pumped current \cite{Wei00}. Using this simple model one can investigate many important 
properties of quantum pumps.

\subsection{Brouwer's adiabatic approach}

We consider 1D two harmonically oscillating $\delta$-function barriers with the strengthes 
$X_1=V_1+\lambda_1 \cos \omega \tau$ and $X_2=V_2+\lambda_2 \cos (\omega \tau + \phi)$
respectively, separated by a distance $d$. This is a simplified model of the
experiment by Switkes et al., but possesses many important characteristics and can be easily
handled. Wei et al. studied parametric charge pumping aided by quantum resonance using this model
and found that the pumped current has large values near a resonance level \cite{Wei00}. Due to 
the double barrier geometry, resonant tunneling also plays an important role in charge pumping.
This is not so surprising because the adiabatically pumped charge is proportional to the variation
of the scattering matrix to the external parameters. Usually the elements of a scattering matrix are
very senstive to the variation of the external parameters near the resonance.

The $2 \times 2$ scattering matrix of the double barriers with the strengthes $X_1$ and $X_2$ is given by
\begin{equation}
S=\left( 
\begin{array}{cc}
r & t' \\
t & r'
\end{array}
\right),
\end{equation}
where $r$ and $t$ are the reflection and the transmission amplitudes respectively, for modes
incident from the left; $r'$ and $t'$ are similar quantities for modes incident from the right.
The charge emitted per cycle to the right is obtained from
\begin{equation}
Q_1 = e \int_0^{T_p} d\tau \left( \frac{dn_1}{dX_1}\frac{dX_1}{d\tau} 
+ \frac{dn_1}{dX_2}\frac{dX_2}{d\tau} \right),
\label{Brouwer_charge}
\end{equation}
where 
\begin{equation}
\frac{dn_1}{dX_m} = {\rm Im} 
\left( \frac{\partial t}{\partial X_m}t^* + \frac{\partial r'}{\partial X_m}r'^* \right)
\end{equation}
($m=1,2$), and $T_p (=2\pi/\omega)$ is the period of the pumping. One can show that the emitted 
charge to the left $Q_2$ is equal to $-Q_1$. Equation (\ref{Brouwer_charge}) can be rewritten 
in the following form by using Green's theorem
\begin{equation}
Q_1 = e \int_A dX_1 dX_2 \Pi (X_1,X_2),
\label{Brouwer_surface}
\end{equation}
where $\Pi (X_1,X_2) = \partial (dn_1/dX_2)/\partial X_1 - \partial (dn_1/dX_1)/\partial X_2$.
The pumped current is easily obtained from $I_1 = Q_1/T_p$. 

\begin{figure}
\center
\includegraphics[height=\size cm,angle=0]{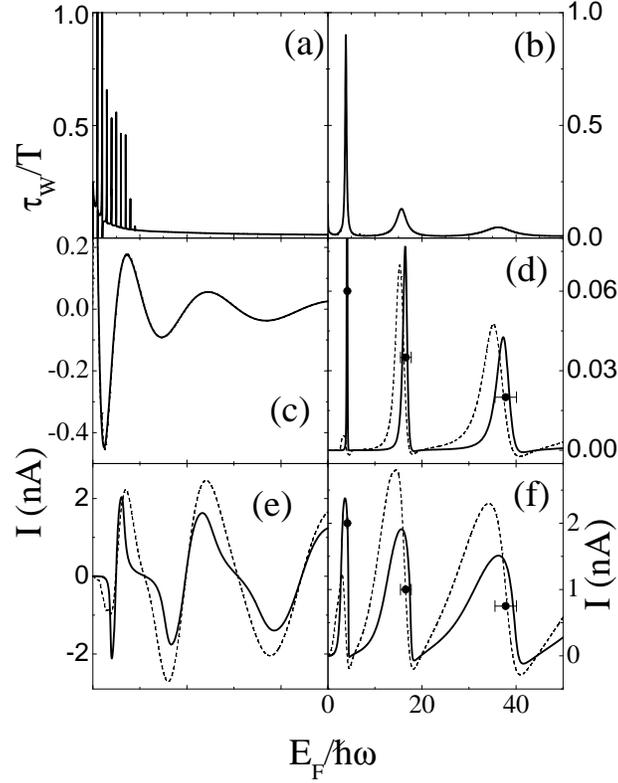}
\caption{The Wigner delay times (a) with the same condition as (c), and (b) as (d).
The pumped current $I_1$ calculated by using Brouwer's approach (solid curves) 
and the Floquet approach (dotted curves) with $\phi=\pi/2$ for (c) $\lambda = 22.5$ 
meV$\cdot$nm and (e) $\lambda = 225$ meV$\cdot$nm with $V=0$, and (d) $\lambda = 22.5$ 
meV$\cdot$nm and (f) $\lambda = 225$ meV$\cdot$nm with $V=225$ meV$\cdot$nm. In (d) and (f) 
the transmission resonances are denoted by the filled circles ($\bullet$) obtained from 
considering static double barriers, whose $y$ values are chosen arbitrarily. The attached 
error bars represent the sizes of the imaginary energy of each resonance.}
\label{fig4}
\end{figure}

We use the parameters $d=50$ nm for the distance between the two barriers, the effective mass 
$\mu = 0.067m_e$ of an electron in GaAs, and $T_p=9.09$ ps for the period of pumping, which 
corresponds to $\hbar \omega = 0.45$ meV. Figure \ref{fig4} shows the pumped current as a 
function of the energy of an incident electron with $V_1 = V_2 = V$, $\lambda_1 = \lambda_2 = 
\lambda$, and $\phi=\pi/2$. We present two examples; the first is a nearly open case ($V$=0)
in Figs.~\ref{fig4}c and \ref{fig4}e, and the second is a closed case ($V$=225 meV$\cdot$nm)
in Figs.~\ref{fig4}d and \ref{fig4}f. All of them show interesting resonance-like structures
(See Ref. \cite{Kim02} for the details of the resonance-like behavior).

\subsection{Floquet approach}

Now we study the Floquet approach of the problem investigated in Sec.~II. Using the scattering 
matrix of a single $\delta$-function with sinusoidal time dependence (see Appendix for details),
we can obtain the total scattering matrix of the oscillating double $\delta$-functions in the 
following form 
\begin{equation}
S=\left( 
\begin{array}{cc}
{\bf r} & {\bf t'}\\
{\bf t} & {\bf r'} 
\end{array}
\right),
\end{equation}
where ${\bf r} = {\bf r}_L + {\bf t}_L({\bf I}-{\bf Qr}_R{\bf Qr}_L)^{-1}{\bf Qr}_R{\bf Qt}_L$, and 
${\bf t} = {\bf t}_R({\bf I}-{\bf Qr}_L{\bf Qr}_R)^{-1}{\bf Qt}_L$; ${\bf r}'$ and ${\bf t}'$ 
can also be obtained by replacing $L$ by $R$ in ${\bf r}$ and ${\bf t}$ respectively. Here 
${\bf r}_{R(L)}$ and ${\bf t}_{R(L)}$ are the reflection and the transmission matrices respectively, 
for the right (left) delta function with time dependence for modes from the left, and ${\bf I}$ is 
an identity matrix. Due to the reflection symmetry of each delta function 
${\bf r}'_{R(L)} = {\bf r}_{R(L)}$, and ${\bf t}'_{R(L)} = {\bf t}_{R(L)}$. During each one-way trip 
an electron at energy $E_n = E +n\hbar\omega$ picks up a phase factor $\exp(ik_nd)$, which is 
represented by the diagonal matrix ${\bf Q}_{mn}=\exp(ik_md)\delta_{mn}$. Once we have Floquet
scattering matrix $S$, the pumped current to the right is given by Eq.~(\ref{final}), i.e.
\begin{equation}
I = \frac{e}{h} \int dE dE' [t(E',E) f_L(E) - t'(E,E') f_R(E')].
\label{datta_current}
\end{equation}
Without external bias ($f_L = f_R = f$) Eq.~(\ref{datta_current}) can be rewritten as
\begin{equation}
I = \frac{2e}{h} \int_0^\infty dE f(E) [T^+(E)-T^-(E)],
\label{current}
\end{equation}
where $T^+(E)=\sum_{E_n>0}T^+_{En,E}$. At zero temperature it becomes
\begin{equation}
I = \frac{2e}{h}\int_0^{E_F} dE [T^+(E)-T^-(E)],
\label{current_Fl}
\end{equation}
where $E_F$ is the Fermi energy. 

Figure \ref{fig4} shows that in the open case ($V=0$) for small $\lambda$ (Fig.~\ref{fig4}c)
the pumped current obtained from the Floquet approach is equivalent to that of Brouwer's while 
for rather larger $\lambda$ (Fig.~\ref{fig4}e) they deviate from each other, which is not
fully understood yet. When $V \neq 0$ (the closed case), even for small $\lambda$ 
(Fig.~\ref{fig4}d) they are quantitatively different near the resonances. It is answered in 
the following subsection.

\subsection{Wigner delay time}

Figures \ref{fig4}a and b show the Wigner delay times by using the same parameters exploited in 
Fig.~\ref{fig4}c and d, respectively. In Fig.~\ref{fig4}a the Wigner delay time decreases as $E$ increases, 
which can be understood when one takes it into account that usually the electron dwell time is short 
if the energy of an incident electron is large. Note that the sharp spikes for low energies are just artifacts 
coming from the fact that the Floquet scattering matrix numerically obtained diverges at $E=n\hbar\omega$.
Since the Wigner delay time is small enough, the adiabatic condition is still fulfiled to show very
good agreement between the two currents in Fig.~\ref{fig4}c.
In Fig.~\ref{fig4}b several peaks of the Wigner delay time is clearly shown. It can be explained by
thinking that now two static barriers are plugged into the problem, which generates the so-called
double barrier resonances. The Wigner delay times have large values near the resonances since 
at the resonances an electron can stay in the quantum dot for a long time, which immedeatly implies that 
the adiabatic condition can be broken down. It explains the deviation between the pumped currents 
near the resonances observed in Fig.~\ref{fig4}d.

\begin{figure}
\center
\includegraphics[height=12 cm,angle=-90]{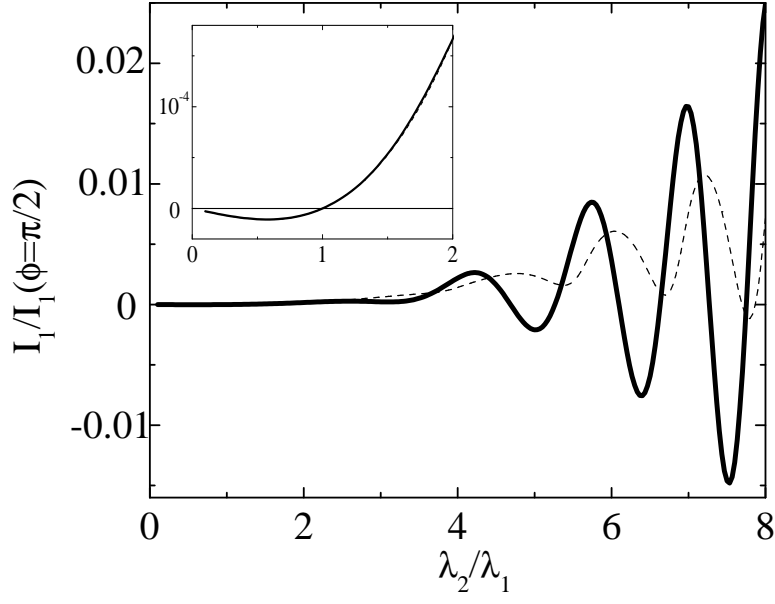}
\caption{Pumped current $I_1$ as a function of potential asymmetry $\lambda_2/\lambda_1$
with $\lambda_1=22.5$ meV$\cdot$nm and $E=6.005\hbar\omega$ at $\phi=0$ (solid curve) and
$\phi=\pi$ (dashed curve), where the currents are normalized to their values at at $\phi=\pi/2$.
Inset shows the magnification of a part of the plot.}
\label{fig5}
\end{figure}

\subsection{Broken symmetry}

One of the interesting consequences from Eq.~(\ref{current_Fl}) is that the pumped current still exists
even in the cases $\phi=0$ or $\pi$ when $\lambda_1 \neq \lambda_2$ \cite{Zhu02}. Since the integration 
area in parameter space is zero when $\phi=0$ or $\pi$, in Brouwer's approach the pumped current 
definitely vanishes. In contrast, even when $\phi=0$ or $\pi$, an asymmetry of the potential can 
lead to the asymmetry of the currents \cite{Datta92}, which is nothing but the pumped current in 
Eq.~(\ref{current_Fl}). Figure \ref{fig5} shows the pumed current as a function of ratio of the 
strength of two barriers $\lambda_2/\lambda_1$ with $\phi=0$ or $\pi$, and $E=6.005\hbar\omega$. 
Note that the pumped current is zero when $\lambda_1 = \lambda_2$. The oscillatory behavior is also 
related to the double barrier resonances.

\subsection{Pauli blocking factor}

\begin{figure}
\center
\includegraphics[height=8 cm,angle=0]{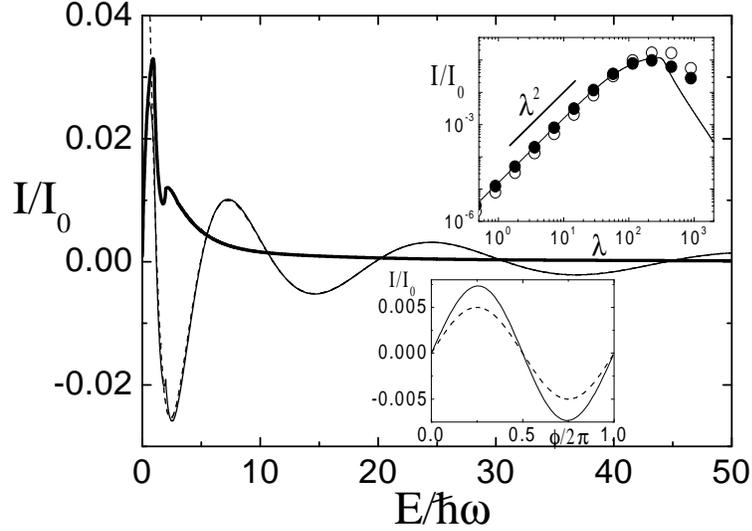}
\caption{The pumped currents obtained from the Floquet approach with (the thick solid
curve) and without Pauli blocking (the thin solid curve), and Brouwer's formula 
(the dashed curve) with the same parameters used in Fig.~\ref{fig1}.
Note that $I_0=e/T=17.6$ nA.
The upper inset: the pumped currents as a function of the strength of the oscillating
potential $\lambda$ from the Floquet approach with (the filled circles) and without Pauli 
blocking (the open circles), and Brouwer's formula (the solid curve). The lower inset:
the pumped currents as a function of a phase difference $\phi$ from the Floquet approach 
with (the solid curve) and without Pauli blocking (the dashed curve). In both insets
we take $E=6.005\hbar\omega$}
\label{fig6}
\end{figure}

Figure \ref{fig6} shows the pumped currents obtained from Floquet approach with and without Pauli blocking 
factor, and Brouwer's formula in the same model used above under the adiabatic regime as shown in 
Fig.~\ref{fig4}c. It is clearly seen that the current with Pauli blocking factor deviates from that of Brouwer's 
approach which nearly coincide with the current without Pauli blocking factor. It is worth nothing that qualitative 
behavior of the pumped currents with and without Pauli blocking looks quite similar: the pumped current 
$I \propto \lambda^2\sin\phi$ as shown in the insets of Fig.~\ref{fig6}. 

\section{Summary and Discussion}

In summary, we have reviewed Floquet formalism of quantum electron pumps and shown its various application. In the 
Floquet formalism the quantum pump is regarded as a time dependent scattering system, which allows us to go beyond 
the adiabatic limit. We have derived the expression of the pumped current using scattering states, and shown that the
adiabatic formula given by Brouwer is successfully recovered in the limit $\omega \rightarrow 0$. The condition
for obtaining non-zero pumped current is rather subtle when the non-adiabatic case is considered. In the adiabatic
case the time reversal symmetry should be broken to have the pumped current, while it is sufficient to generate the
pumped current to break either the spatial reflection symmetry or the time reversal symmetry in the non-adiabatic
case. Since usually the quantum pump can break the time reversal symmetry with time dependent perturbations,
the issue of Pauli blocking originating from the Fermionic nature of electrons can occur. It can be shown, however, 
that the Pauli blocking factor is not neccessary at all. Using the Floquet formalism one can discuss the symmetry 
properties of the pumped current over the inversion of the static magnetic field for various spatial symmetries of 
quantum dots. Application of the Floquet formalism dierctly to realistic problems such as 2D quantum dots is rather 
difficult because one should consider many sidebands for each scattering channel. However, we can get many 
basic important physics of quantum pumps by studying a rather simple model system: a 1D oscillating double 
$\delta$-function barriers.

The adiabatic charge pumping does not have any corresponding classical analog since it has something to do
with Berry phase, which means this is completely quantum mechanical phenomena. The quantum pump in the 
non-adiabatic regime, however, is somewhat subtle because Berry phase concept dose not work any longer.
It has been shown that there are several other mechanisms to generate dc current through the non-adiabatic 
time periodic perturbations with zero average, e.g. quantum ratchets \cite{Reimann97,Linke99,Astumian02}, 
quantum ballistic rectifiers \cite{Song98,Fleischmann02}, Hamiltonain ratchet \cite{Schanz01}, and quantum 
shuttles \cite{Gorelik98,Park00}.
We believe that it must be important to understand the relations and the differences among these devices.
The Flouet formalism might play an important role to understand them because it allows us to go beyond the adiabatic
limit of quantum pumps. Finally, we would like to mention that it is also important to know what role the decoherence
plays in the system with time periodic perturbation. It helps us understand the classical-quantum correspondence 
of dc current generation of quantum pumps, ratchets, rectifiers, and shuttles.

The Floquet formalism has been successively used to describe a lot of physical properties of the quantum pump.
It is a powerful tool not only for understanding quantum pumps, which is the main topic of this review, but also
for many kinds of coherent transport problems with time-periodically varying scatterers.


\section*{Acknowledgements}
This work was supported by Research Institute for Basic Sciences, Pusan National University (2004), and 
Pusan National University Research Grant, 2004.


\appendix

\section{Floquet scattering matrix in a single oscillating $\delta$-function impurity}

The scattering problem of a single $\delta$-function impurity with sinusoidal time dependence has been 
investigated by several authors \cite{Bagwell92,Martinez01,Kim02}. We would like to summarize how to 
construct its Floquet scattering matrix in this Appendix. The system is described by the Hamiltonian
\begin{equation}
H(x,t) = -\frac{\hbar^2}{2\mu}\frac{d^2}{dx^2} + [V_s + V_d \cos(\omega t + \phi)]\delta(x),
\end{equation}
where $\mu$ is the mass of the incident particle, while $V_s$ and $V_d$ represent the
strength of the static and the oscillating potential respectively. Using the Floquet
formalism the solution of this Hamiltonian can be expressed as
\begin{equation}
\Psi_{E_{Fl}}(x,t) = e^{-i E_{Fl} t/\hbar}\sum_{n=-\infty}^{\infty} \psi_n (x)
e^{-in\omega t},
\label{floquet_ap}
\end{equation}
where $E_{Fl}$ is the Floquet energy which take continuous values in the interval
$0 < E_{Fl} \leq \hbar \omega$.

Since the potential is zero everywhere except at $x=0$, $\psi_n(x)$ is given by the
following form
\begin{equation}
\psi_n (x) = \left\{
\begin{array}{c}
A_n e^{ik_n x} + B_n e^{-ik_n x}, ~~~ x<0 \\
C_n e^{ik_n x} + D_n e^{-ik_n x}, ~~~ x>0,
\end{array} \right.
\label{plane_wave_ap}
\end{equation}
where $k_n=\sqrt{2\mu(E_{Fl} + n\hbar\omega)}/\hbar$.
The wave function
$\Psi_{E_{Fl}}(x,t)$ is continuous at $x=0$,
\begin{equation}
A_n + B_n = C_n + D_n,
\label{eq1.1}
\label{osc_bc1}
\end{equation}
and the derivative jumps by
\begin{equation}
\left.\frac{d\Psi_{E_{Fl}}}{dx}\right|_{x=0^+} - \left.\frac{d\Psi_{E_{Fl}}}{dx}\right|_{x=0^-}
=\frac{2m}{\hbar^2}[V_s + V_d \cos(\omega t + \phi)]\Psi_{E_{Fl}}(0,t).
\label{osc_bc2}
\end{equation}
Using Eq.~(\ref{floquet_ap}) this leads to the condition
\begin{eqnarray}
&& ik_n (C_n - D_n - A_n + B_n) \nonumber \\
&=& \gamma_s(A_n+B_n) +\gamma_d(e^{-i\phi}A_{n+1}+e^{i\phi}A_{n-1}+e^{-i\phi}B_{n+1}+e^{i\phi}B_{n-1}) \label{eq2} \\
&=& \gamma_s(C_n+D_n) +\gamma_d(e^{-i\phi}C_{n+1}+e^{i\phi}C_{n-1}+e^{-i\phi}D_{n+1}+e^{i\phi}D_{n-1}) \nonumber
,\end{eqnarray}
where $\gamma_s=2\mu V_s/\hbar^2$ and $\gamma_d=\mu V_d/\hbar^2$. After some algebra
we have the following equation from Eqs.~(\ref{eq1.1}) and (\ref{eq2})
\begin{equation}
\left(
        \begin{array}{c}
        \vec{B} \\ \vec{C}
        \end{array}
\right)
=
\left(
        \begin{array}{cc}
        -(I+\Gamma)^{-1}\Gamma & (I+\Gamma)^{-1} \\
        (I+\Gamma)^{-1} & -(I+\Gamma)^{-1}\Gamma
        \end{array}
\right)
\left(
        \begin{array}{c}
        \vec{A} \\ \vec{D}
        \end{array}
\right)
\label{mat_eq_ap}
,\end{equation}
where
\begin{equation}
\Gamma =
\left(
        \begin{array}{ccccc}
        \ddots & \ddots & 0 & 0 & 0 \\
        \gamma_d e^{i\phi}/ik_{-1} & \gamma_s/ik_{-1} & \gamma_d e^{-i\phi}/ik_{-1} & 0 & 0 \\
        0 & \gamma_d e^{i\phi}/ik_0 & \gamma_s/ik_0 & \gamma_d e^{-i\phi}/ik_0 & 0 \\
        0 & 0 & \gamma_d e^{i\phi}/ik_1 & \gamma_s/ik_1 & \gamma_d e^{-i\phi}/ik_1\\
        0 & 0 & 0 & \ddots & \ddots
        \end{array}
\right),
\label{mat_eq2_ap}
\end{equation}
and $I$ is an infinite-dimensional square identity matrix. Eq.~(\ref{mat_eq_ap}) can also be expressed in the
form $\left.|{\rm out}\right> = M \left.|{\rm in}\right>$, where $M$ connects the input coefficients to 
the output coefficients including the associated evanescent Floquet sidebands. In order to construct the 
scattering matrix we multiply an identity to both sides, 
$K^{-1}K\left.|{\rm out}\right> = M K^{-1}K\left.|{\rm in}\right>$, where $K_{nm}=\sqrt{k_n}
\delta_{nm}$. Then we have $\vec{J}_{out}=\bar{M}\vec{J}_{in}$, where $\vec{J}$ represents
the amplitude of probability flux and $\bar{M} \equiv KMK^{-1}$. It should be mentioned
that $\bar{M}$ is not unitary due to the evanescent modes included.
If we keep only the propagating modes, we obtain the unitary scattering matrix $S$.



\begin{thebibliography}{0}
\bibitem{cal_detail}
Due to $\nabla$ one can easily find that 
$\overline{\chi^{\sigma'*}_{E'} \nabla \chi^\sigma_E} \propto \delta_{\sigma'\sigma} 
\delta(E-E')$, where $\overline{(\cdots)}$ represents space and time average at
$x\rightarrow +\infty$.
\end{thebibliography}
\end{document}